\documentclass[showkeys,superscriptaddress,pra,aps,12pt]{revtex4-1}
\usepackage[utf8]{inputenc}
\usepackage{graphicx}
\usepackage{amssymb}
\usepackage{amsmath}
\usepackage{mathtools}
\usepackage{comment}
\usepackage{braket}
\usepackage{xcolor}

\DeclareMathOperator{\Tr}{Tr}

\begin{document}

\title{Star Product Formalism for Probability and Mean Value Representations of Qudits}

\author{Peter Adam}
\email{adam.peter@wigner.mta.hu}
\affiliation{Institute for Solid State Physics and Optics, Wigner Research Centre for Physics, H-1525 Budapest, P.O. Box 49, Hungary}
\affiliation{Institute of Physics, University of P\'ecs, Ifj\'us\'ag \'utja 6, H-7624 P\'ecs, Hungary}
\author{Vladimir A. Andreev}
\affiliation{Lebedev Physical Institute, Russian Academy of Sciences, Leninskii Prospect 53, 119991 Moscow, Russia}
\author{Margarita A. Man'ko}
\affiliation{Lebedev Physical Institute, Russian Academy of Sciences, Leninskii Prospect 53, 119991 Moscow, Russia}
\author{Matyas Mechler}
\affiliation{Institute of Physics, University of P\'ecs, Ifj\'us\'ag \'utja 6, H-7624 P\'ecs, Hungary}
\author{Vladimir I. Man'ko}
\affiliation{Lebedev Physical Institute, Russian Academy of Sciences, Leninskii Prospect 53, 119991 Moscow, Russia}

\begin{abstract}
The quantizer--dequantizer formalism is developed for mean value and probability representation of qubits and qutrits. We derive the star-product kernels providing the possibility to derive explicit expressions of the associative product of the symbols of the density operators and quantum observables for qubits.
We discuss an extension of the quantizer--dequantizer formalism associated with the probability and observable mean-value descriptions of quantum states for qudits.
\end{abstract}
\maketitle

\section{Introduction}

The phase-space formulation of quantum mechanics is still in the focus of research interest, due to its numerous important applications \cite{Ferrie2011, Filippov2012, MessamahQIP2015, Garon2015, Romero2015, CamposAnnPhys2018}.
The states of a quantum system can be completely described by quasiprobability distributions such as the Wigner function \cite{Wigner32}, the Husimi Q-function \cite{Husimi40,Kano46} and the Glauber-Sudarshan P-function \cite{Glauber63,Sudarshan63} defined in phase space. These functions are widely used for calculations in various physical problems, especially in quantum optics \cite{HOL,HSW,KC,MW,SZ,WS}.
A fair probability distribution called the symplectic tomogram has also been introduced in connection with measuring the quantum states of light by means of optical homodyne tomography \cite{ManciniPLA96,IbortPS2009,AmosovPRA2012,MarmoPS2015}.

The idea of the phase-space formalism can be extended to finite dimensional quantum systems used in quantum information processing. Finding either a complete, continuous Wigner function \cite{Dowling1994, Kastrup2016, Tilma2016} or discrete Wigner functions having the same essential properties as their continuous counterparts \cite{MunozQIP2017, WKW, UL2, AB, GHW, CMM, KRS, ERL, FM2} for such systems is still a subject of investigations.
Tomographic probability distributions called spin tomograms \cite{DodPLA,OlgaJETP,AndreevJETP,AMMS}, and unitary matrix tomograms \cite{MMSV} have also been developed for finite dimensional spin systems.

In order to use quasi-probability distributions and tomograms in physical problems the operators modeling observable physical quantities have to be represented \cite{Stratonovich}. This representation is called the symbol of operators. The algebra of symbols corresponding all possible manipulations with operators on the Hilbert space can be constructed by applying the general star-product scheme \cite{FFL,Brif1999,OlgaMarmoJPA}. Within this formalism one can relate operators to their symbols using dequantizers and can reconstruct operators from their symbols using quantizers \cite{LizziVitale}. The product of operators is mapped onto an associative product of symbols of operators called star product defined by an integral containing an integral kernel. The kernel can be derived using quantizer and dequantizer operators \cite{MarmoPS2015,Marmo-JPA2017,Marmo-FP2017,NuovoCimento,IMMSVS}. The relations between different phase-space representations can be also determined in this framework \cite{OlgaMarmoJPA,Vourdas2006,ADM1,ADM2}.

The star product formalism of symbols for $N$-dimensional systems is described in detail in \cite {AAMM}. For qubit states, the set of quantizers and dequantizers for the spin tomogram was considered, e.g., in~\cite{Filippov} and a detailed analysis of the spin Wigner functions and probability distributions is given in~\cite{AdamPLA,JRLR-6-2017,JRLR-4-2018}. Using this formalism the relations between tomograms and Wigner functions for one and two qubits have been determined  \cite{AAM1,AAM2,AAMM}.

Recently, a specific probability description of qubit and qutrit states was introduced in~\cite{ChernegaJRLR17a,ChernegaJRLR17b,ChernegaJRLR17c}, where the approach was called quantum suprematism. In this representation the density matrix elements for qubits are expressed by three probabilities to have spin projections $m=+1/2$ onto the $x$, $y$ and $z$ axis, while for qutrits the probabilities are measured on three artificial qubits composed of the three basis states of the qutrit.
The states of qubits and qutrits can be mapped onto geometrical objects -- triads of squares on a plane, called the \textit{Triad of Malevich's squares}~\cite{Shatskikh}. The areas of the squares obey to the quantum constraint expressed in terms of inequalities for the probabilities determining the qubit and qutrit density matrices. In another general form of the qubit state density matrix the matrix elements are expressed by the mean values of the spin projections onto the $x$, $y$, and $z$ axis. A major advantage of these representations is that the symbols of the density operators are measurable quantities.

The aim of this study is to derive the quantizer-dequantizer formalism for the probability and mean value representations of qubit and qutrit density matrices.
We find in an explicit form the dequantizers that create the probability and mean value symbols and the corresponding quantizers that provide the reconstruction of the density matrices. In the case of qubits we also determine the kernel matrices which are required for the derivation of the star product of the symbols. Using these kernels we derive the structure constants of the Lie algebra of the quantizer operators. We discuss how the formalism can be extended to qudit states.

This paper is organized as follows.

We review a generic scheme of quantizers and dequantizers in section~\ref{sec:FQaD}. In sections~\ref{sec:QbS} and \ref{sec:SoQtQaD}, we derive the quantizer and dequantizer operators for probability and mean value representation of qubits and qutrits, respectively. In section \ref{sec:KASPDQD} we present the star-product formalism for the probability and mean value representations of qubits and the necessary kernel matrices are derived.  Finally,  we conclude in section \ref{sec:conc}.

\section{Formalism of quantizers and dequantizers}\label{sec:FQaD}

In this section we summarize briefly the general formalism of using $c$-number functions instead of operators to describe quantum systems.

All invertible maps connecting operators acting in the Hilbert space $\cal H$ and functions of some variables can be described using two families of operators called  dequantizers $\hat U(x)$ and quantizers $\hat D(x)$~\cite{OlgaMarmoJPA}. Parameters $x$ label points in a manifold; they can be either continuous or discrete.

One can construct a  $c$-number function $f_A(x)$, called the symbol of the operator $\hat A$ using the definition
\begin{equation}\label{F1}
f_A(x)=\Tr\left(\hat A\hat U(x)\right).
\end{equation}
The operator $\hat A$ can be expressed in terms of the symbol $f_A(x)$ of the operator as
\begin{equation}\label{F2}
\hat A=\int f_A(x)\hat D_A(x)\,dx.
\end{equation}
Multiplying Eq.~\eqref{F2} by the operator $\hat{U}(x')$ and taking the trace we get
\begin{equation}
\label{eq:2:3}
f_A(x')=\int f_A(x) \Tr \left(\hat D(x)\hat U(x')\right)dx
\end{equation}
and consequently we obtain a consistency condition
\begin{equation}\label{F4}
\Tr\left(\hat D(x)\hat U(x')\right)=\delta(x-x').
\end{equation}
 for the operators $\hat U(x')$ and $\hat D(x)$.

In this formalism, the product of operators $\hat A$ and $\hat B$ is mapped onto the product of symbols of operators $(f_A\star f_B)(x)$ called star product and defined by
\begin{equation}\label{F5}
(f_A\star f_B)(x_3)=\int f_A(x_1)f_B(x_2)K(x_1,x_2x_3)\,dx_1\,dx_2,
\end{equation}
where the kernel of the product is
\begin{equation}\label{F6}
K(x_1,x_2,x_3)=\Tr\left(\hat D(x_1)\hat D(x_2)\hat U(x_3)\right).
\end{equation}
If parameters $x$ are discrete, the integrals in the above formulas
are replaced by sums over discrete parameters $x$ and the term $\delta(x-x')$
changes to Kronecker index $\delta_{xx'}$. For example, formulas \eqref{F1}, \eqref{F2} and \eqref{F4} are modified as
\begin{equation}\label{F1d}
f_A^{(i)}=\Tr\left(\hat A\hat U^{(i)}\right),\qquad i=1,\ldots,n,
\end{equation}
\begin{equation}\label{F2d}
\hat A=\sum_{i=1}^n f_A^{(i)}\hat D_A^{(i)},
\end{equation}
\begin{equation}\label{F4d}
\Tr\left(\hat D^{(i)}\hat U^{(j)}\right)=\delta_{ij}, \qquad i,j=1,\ldots, n.
\end{equation}
For self-dual sets the set of quantizers coincides with the corresponding set of dequantizers, that is, $\hat D^{(i)}=\hat U^{(i)},\, i=1,\ldots,n$, so self-dual dequantizer operators satisfy the orthogonality condition
\begin{equation}
\label{F22} \Tr\left(\hat U^{(i)} \hat U^{(j)}\right)=\delta_{ij},\qquad i,j=1,\ldots,n.
\end{equation}

In our study, we consider the operators for qubit and qutrit systems. In the above equation, for qubits, $n=4$, for qutrits, $n=9$, while for a $d$-dimensional qudit quantum system, $n=d^2$. In view of the possible physical applications, important issue is the physical meaning of the symbols of the operators. 

\section{Quantizers and dequantizers for probability and mean value representation of qubits}\label{sec:QbS}

A density matrix $\rho$ of a qubit can be expressed within the framework of the probability representation~\cite{Marmo-JPA2017,ChernegaJRLR17a,ChernegaJRLR17b,ChernegaJRLR17c}  in terms of three probabilities $0\leq p_1,p_2,p_3\leq 1$ as
\begin{equation}\label{F15}
\rho=\begin{pmatrix}
p_3&(p_1-1/2)-i(p_2-1/2)\\
(p_1-1/2)+i(p_2-1/2)&1-p_3
\end{pmatrix},
\end{equation}
where $p_1$, $p_2$, and $p_3$ are the probabilities to have spin projections
$m=+1/2$ onto the $x$, $y$, and $z$ axes, respectively.
In view of the nonnegativity of the density matrix $\rho$, the probabilities satisfy the constraint
\begin{equation}\label{F16}
(p_1-1/2)^2+(p_2-1/2)^2+(p_3-1/2)^2\leq 1/4.
\end{equation}

\begin{figure}
\centerline{\includegraphics[width=0.5\columnwidth]{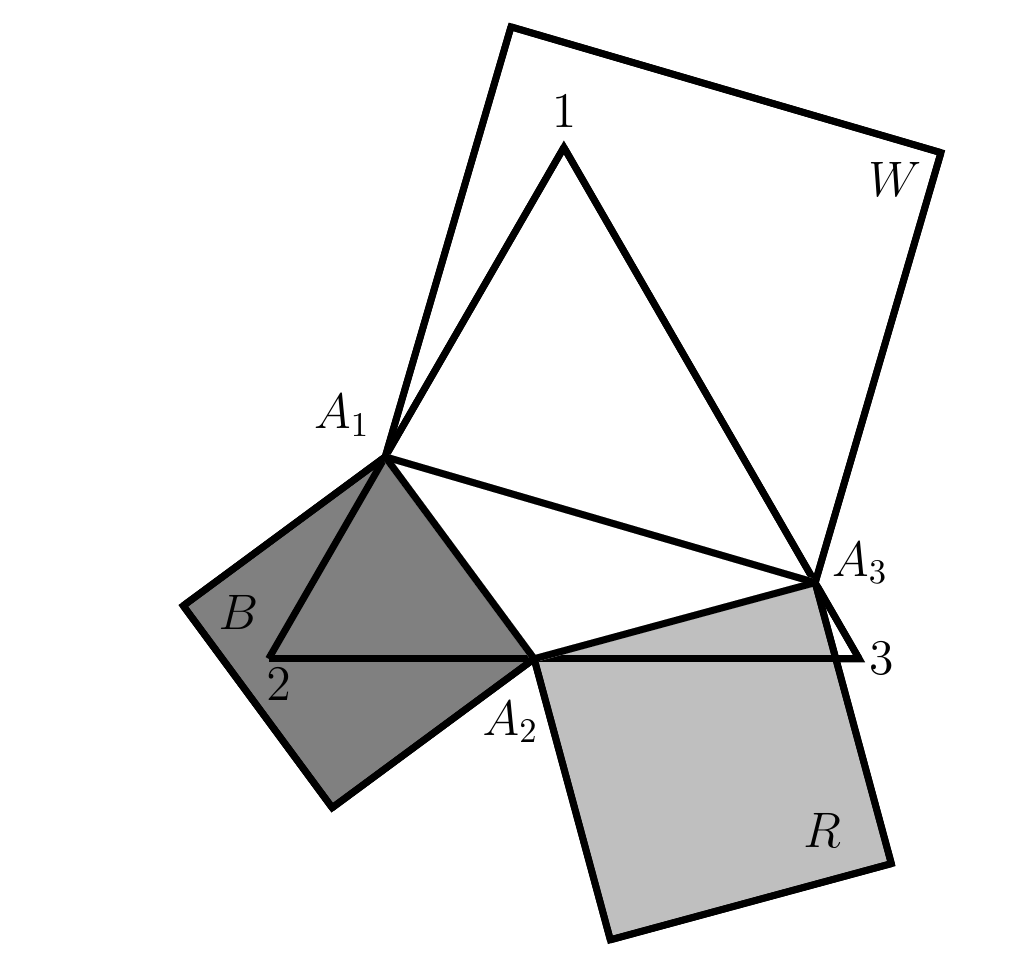}}
\caption{ \textit{Triad of Malevich's squares} illustrating the qubit-state density matrix.\label{fig:TMS}}
\end{figure}

The state with the density matrix~\eqref{F15} can be illustrated by the \textit{Triad of Malevich's squares} shown in Fig.~\ref{fig:TMS} \cite{ChernegaJRLR17a,ChernegaJRLR17b,ChernegaJRLR17c,CMM1,CMM2}. In this picture the equilateral triangle $123$ has sides with length $\sqrt 2$ and the triangle $A_1A_2A_3$ is determined by the probabilities $p_1$, $p_2$, and $p_3$.
Malevich's squares (black - $B$, red - $R$, and white - $W$) are constructed using the sides of the triangle $A_1A_2A_3$. The sum of areas of the squares is expressed in terms of the probabilities $p_1$, $p_2$, and $p_3$ as
follows~\cite{ChernegaJRLR17a,ChernegaJRLR17b,ChernegaJRLR17c,CMM1,CMM2,mmanko-JPCS,Milestones,Entropy-MA-2018,
Entropy-Julio}:
\begin{equation}\label{F17}
S=2\left[3(1-p_1-p_2-p_3)+2p_1^2+2p_2^2+2p_3^2+p_1p_2+p_2p_3+p_3p_1\right].
\end{equation}
For qubit state, the condition $\mbox{det}\,\rho\geq 0$ provides
the constraint  $S\leq3$ for the sum \eqref{F17} that can be verified experimentally.

In order to find the dequantizer operators leading to a $c$-number representation in which the probabilities $p_i$, $i=1,2,3$ are the symbols of the operator $\hat\rho$ the following procedure can be applied.
We are looking for Hermitian dequantizer operators defined through the matrices in the general form
\begin{equation}
U^{(i)}=
\begin{pmatrix}
u^{(i)}_1&u^{(i)}_2-iu^{(i)}_3\\
u^{(i)}_2+iu^{(i)}_3&u^{(i)}_4
\end{pmatrix},\qquad i=1,\ldots,4.
\end{equation}
Using Eq.~\eqref{F1d} with $f_i=p_i$, $i=1,2,3$, $f_4=1-p_3$, and $\hat{A}=\hat{\rho}$, we arrive at four linear equations each containing three arbitrary parameters $p_i$, $i=1,2,3$ and sixteen variables. Aiming at determining these sixteen variables, we create the four-by-five augmented coefficient matrix of the system of equations, each element of which must be equal to zero. As a result, we arrive at sixteen equations for sixteen variables that can be easily solved, leading to the following dequantizer matrices:
\begin{eqnarray}
\label{F10} 
\begin{aligned}
U^{(1)}&=\frac12\begin{pmatrix} 1&1\\1&1
\end{pmatrix},\qquad&
 U^{(2)}&=\frac12\begin{pmatrix*}[r] 1&-i\\i&1
\end{pmatrix*},\\
 U^{(3)}&=\begin{pmatrix} 1&0\\0&0
\end{pmatrix},&
 U^{(4)}&=\begin{pmatrix} 0&0\\0&1
\end{pmatrix}.
\end{aligned}
\end{eqnarray}
The corresponding set of quantizer operators can be obtained by applying the orthogonality condition \eqref{F4d} to the dequantizers $U^{(i)}$, $i=1,\ldots,4$ and four generic Hermitian matrices leading to the quantizers represented by the matrices

\begin{eqnarray}
\label{F13}
\begin{aligned}
 D^{(1)}&=\begin{pmatrix} 0&1\\1&0
\end{pmatrix},\qquad&
D^{(2)}&=\begin{pmatrix*}[r] 0&-i\\i&0
\end{pmatrix*},\\
D^{(3)}&=\begin{pmatrix} 1&\dfrac{-1+i}2\\\dfrac{-1-i}2&0
\end{pmatrix},&
D^{(4)}&=\begin{pmatrix} 0&\dfrac{-1+i}2\\ \dfrac{-1-i}2&1
\end{pmatrix}.
\end{aligned}
\end{eqnarray}
Using the dequantizer operators defined in Eq.~\eqref{F10} and the density operator \eqref{F15} one can check that the elements of the symbol of the density operator $\rho$ are really the probabilities, that is,
\begin{eqnarray}
\label{New18}
\begin{aligned}
\Tr(\rho U^{(i)})&=p_i,\quad&i=1,2,3,\qquad
\Tr(\rho U^{(4)})&=p_4=1-p_3.
\end{aligned}
\end{eqnarray}
Using the quantizers \eqref{F13} and the symbol \eqref{New18} one can derive the density operator \eqref{F15} in the form
\begin{eqnarray}
\label{F18}
\begin{aligned}
\rho&=\sum_{i=1}^{4}p_iD^{(i)}
\end{aligned}
\end{eqnarray}
in accordance with Eq.~\eqref{F1d}.

Let us consider another general form of the qubit-state density matrix $\rho$ as
\begin{eqnarray}
\label{F25} \rho=\frac12 \begin{pmatrix} 1+z&x-iy\\x
+iy&1-z\end{pmatrix},\qquad x^2+y^2+z^2\leq 1.
\end{eqnarray}
Here numbers $x$, $y$, and $z$, called the Bloch-sphere parameters, are the mean values of spin projections onto the axes $\vec X$, $\vec Y$, and $\vec Z$, respectively. In the following we choose these mean values as three of the symbols $f_i$ of the operator and we find the corresponding Hermitian dequantizer matrices $U^{(i)}$. $i=1,\ldots,4$. Using Eq.~\eqref{F1d} with $f_1=x$, $f_2=y$,  $f_3=z$, and $\hat{A}=\hat{\rho}$ as defined in Eq.~\eqref{F25}, we arrive at three linear equations each containing three arbitrary parameters $x,y,z$ and twelve variables. To determine these variables, a three-by-four augmented coefficient matrix of the system of equations is created the element of which must be equal to zero. Therefore we arrived at twelve equations containing twelve unknowns. Solving this system of equations we find that the resulting matrices are the Pauli matrices. As these matrices are orthogonal to each other, we may expect them to form a self-dual set provided that a suitable fourth matrix can be found. Note, however, that in order to apply condition \eqref{F22} the matrices must be normalized. Using this condition, we arrive at the self-dual set of dequantizer matrices
\begin{eqnarray}
\label{F21}
\begin{aligned}
U^{(1)}_S&=\frac1{\sqrt2}\begin{pmatrix} 0&1\\1&0
\end{pmatrix},\qquad&
U^{(2)}_S&=\frac1{\sqrt2}\begin{pmatrix*}[r] 0&-i\\i&0
\end{pmatrix*},\\
U^{(3)}_S&=\frac1{\sqrt2}\begin{pmatrix*}[r] 1&0\\0&-1
\end{pmatrix*},&
U^{(4)}_S&=\frac1{\sqrt2}\begin{pmatrix} 1&0\\0&1
\end{pmatrix}.
\end{aligned}
\end{eqnarray}
Applying Eq.~\eqref{F1d} to these dequantizer operators one can arrive at the symbol
\begin{eqnarray}
\label{F26}
s_1=\frac1{\sqrt2}x,\quad 
s_2=\frac1{\sqrt2}y,\quad
s_3=\frac1{\sqrt2}z,\quad
s_4=\frac1{\sqrt{2}}.
\end{eqnarray}

Using the dequantizers in Eq.~\eqref{F21} for the density operator $\hat\rho$ given in its probability representation \eqref{F15} the symbol turn out to be
\begin{eqnarray}
s_i=\sqrt2(p_i-1/2),\quad i=1,2,3,\qquad
s_4=1/\sqrt{2}.
\label{F24}
\end{eqnarray}
This symbol can be called mean-value symbol due to the properties described below Eq.~\eqref{F25}.
As we mentioned earlier, the dequantizers in Eq.~\eqref{F21} satisfy the condition~\eqref{F22}, therefore this set of dequantizers form a self-dual set, that is, $D^{(i)}_S=U^{(i)}_S$, $i=1,\ldots,4$. Substituting the quantizers \eqref{F21} and the symbol \eqref{F24} into Eq.~\eqref{F2d}  the density operator \eqref{F15} can be derived.

\section{Quantizers and dequantizers for probability and mean value representation of qutrits}\label{sec:SoQtQaD}
The probability representation of the density operator of a qutrit was introduced in Ref.~\cite{CMM1} in the form
\begin{small}
\begin{eqnarray}
\rho= \begin{pmatrix} p_3^{(33)}+p_3^{(22)}-1&
\left(p_1^{(21)}-\frac{1}{2}\right)-i\left(p_2^{(21)}-\frac{1}{2}\right)&
\left(p_1^{(31)}-\frac{1}{2}\right)-i\left(p_2^{(31)}-\frac{1}{2}\right)\\
\left(p_1^{(21)}-\frac{1}{2}\right)+i\left(p_2^{(21)}-\frac{1}{2}\right)&
1-p_3^{(22)}&
\left(p_1^{(32)}-\frac{1}{2}\right)-i\left(p_2^{(32)}-\frac{1}{2}\right)\\
\left(p_1^{(31)}-\frac{1}{2}\right)+i\left(p_2^{(31)}-\frac{1}{2}\right)&
\left(p_1^{(32)}-\frac{1}{2}\right)+i\left(p_2^{(32)}-\frac{1}{2}\right)&
1-p_3^{(33)}
\end{pmatrix}\nonumber\\\label{F35}
\end{eqnarray}
\end{small}
In the above equation $p_{1,2}^{(jk)}$ ($j,k=1,2,3$, $j>k$) and $p_3^{(jj)}$ ($j=2,3$) are the probabilities of spin-$1/2$ projections equal to $+1/2$ on axes $x$, $y$, $z$, respectively, measured on three artificial qubits composed of the states $\ket{j}$ and $\ket{k}$ that are $\ket{2}$ and $\ket{1}$, $\ket{3}$ and $\ket{1}$, and $\ket{3}$ and $\ket{2}$, respectively, where $\ket{1}$, $\ket{2}$, and $\ket{3}$ are the basis states of the qutrit, e.g.~the states of a three-level atom. The probabilities $p_3^{(jj)}$ are measured on artificial qubits where $k=1$. The qubit representation of qudit states is discussed in Ref.~\cite{LopezQIP2019}.

The dequantizer operators for the probability representation of qutrits can be derived in a similar manner as in the case of qubits. First, we choose the symbols of the operator $\hat\rho$ to be the probabilities used in \eqref{F35}
\begin{eqnarray}
\label{Q3}
\begin{aligned}
f_\rho^{(1)}&= p_1^{(31)},\quad&
f_\rho^{(2)}&= p_2^{(31)},\quad&
f_\rho^{(3)}&= p_3^{(33)},\\
f_\rho^{(4)}&= p_1^{(21)},&
f_\rho^{(5)}&= p_2^{(21)},&
f_\rho^{(6)}&= p_3^{(22)},\\
f_\rho^{(7)}&= p_1^{(32)},&
f_\rho^{(8)}&= p_2^{(32)},&
f_\rho^{(9)}&= 1.
\end{aligned}
\end{eqnarray}
As Eq.~\eqref{F35} contains only eight probabilities, the ninth symbol is chosen to be 1. For qutrits, each generic Hermitian matrix comprises of nine real variables, therefore the set of generic Hermitian matrices used in Eqs.~\eqref{F1d} contains 81 variables. Using such Hermitian matrices, Eq.~\eqref{F1d} leads to nine equations the augmented coefficient matrix of which contains 81 elements. Each of these elements must be equal to zero, that is, to determine the 81 variables of the Hermitian matrices we have 81 linear equations leading to the matrices
\begin{eqnarray}
\begin{aligned}
U^{(1)}&= \frac12\begin{pmatrix}1&0&1\\
0&1&0\\1&0&1
\end{pmatrix},&
U^{(2)}&= \frac12\begin{pmatrix*}[r]1&0&i\\
0&1&0\\-i&0&1
\end{pmatrix*},&
U^{(3)}&= \begin{pmatrix}1&0&0\\
0&1&0\\0&0&0
\end{pmatrix},\\
U^{(4)}&= \frac12\begin{pmatrix}1&1&0\\
1&1&0\\0&0&1
\end{pmatrix},&
U^{(5)}&= \frac12\begin{pmatrix*}[r]1&-i&0\\
i&1&0\\0&0&1
\end{pmatrix*},&
U^{(6)}&= \begin{pmatrix}1&0&0\\
0&0&0\\0&0&1
\end{pmatrix},\\
U^{(7)}&= \frac12\begin{pmatrix}1&0&0\\
0&1&1\\0&1&1
\end{pmatrix},&
U^{(8)}&= \frac12\begin{pmatrix*}[r]1&0&0\\
0&1&i\\0&-i&1
\end{pmatrix*},&
U^{(9)}&= \begin{pmatrix}1&0&0\\
0&1&0\\0&0&1
\end{pmatrix}.
\end{aligned}
\label{F36}
\end{eqnarray}
The corresponding set of quantizer operators can be readily obtained by applying the orthogonality condition \eqref{F4d} with the dequantizers $U^{(i)}$, $i=1,\ldots,9$ and nine generic Hermitian matrices, leading to the matrices
\begin{eqnarray}\nonumber
&&\begin{aligned}
D^{(1)}&= \begin{pmatrix}0&0&1\\
0&0&0\\1&0&0
\end{pmatrix},&
D^{(2)}&= \begin{pmatrix}0&0&i\\
0&0&0\\-i&0&0
\end{pmatrix},&
D^{(3)}= \begin{pmatrix}1&0&0\\
0&0&0\\0&0&-1
\end{pmatrix},\\
D^{(4)}&= \begin{pmatrix}0&1&0\\
1&0&0\\0&0&0
\end{pmatrix},&
D^{(5)}&= \begin{pmatrix}0&-i&0\\
i&0&0\\0&0&0
\end{pmatrix},&
D^{(6)}= \begin{pmatrix}1&0&0\\
0&-1&0\\0&0&0
\end{pmatrix}
,\end{aligned}\\
\nonumber &&D^{(7)}= \begin{pmatrix}0&0&0\\
0&0&1\\0&1&0
\end{pmatrix},\;
D^{(8)}= \begin{pmatrix}0&0&0\\
0&0&i\\0&-i&0
\end{pmatrix},\\
&& D^{(9)}= \frac12\begin{pmatrix}-2&-1+i&-1-i\\
-1-i&2&-1-i\\-1+i&-1+i&2
\end{pmatrix}.
\label{F37}
\end{eqnarray}
It can be easily proved that the reconstruction of the density operator \eqref{F35} is possible by substituting the quantizers \eqref{F37} and the symbol \eqref{Q3} into Eq.~\eqref{F2d}.

Let us now consider the mean-value representation for qutrits. Applying a procedure analogous to the qubit case, by using a symbol similar to the one derived in Eq.~\eqref{F24} one can arrive at the self-dual dequantizer matrices
\begin{eqnarray}
\begin{aligned}
U^{(1)}_S&= \frac1{\sqrt2}\begin{pmatrix}0&1&0\\
1&0&0\\0&0&0
\end{pmatrix},&
U^{(2)}_S&= \frac1{\sqrt2}\begin{pmatrix}0&-i&0\\
i&0&0\\0&0&0
\end{pmatrix},&
U^{(3)}_S&= \frac1{\sqrt2}\begin{pmatrix}0&0&1\\
0&0&0\\1&0&0
\end{pmatrix},\\
U^{(4)}_S&= \frac1{\sqrt2}\begin{pmatrix}0&0&i\\
0&0&0\\-i&0&0
\end{pmatrix},&
U^{(5)}_S&= \frac1{\sqrt2}\begin{pmatrix}0&0&0\\
0&0&1\\0&1&0
\end{pmatrix},&
U^{(6)}_S&= \frac1{\sqrt2}\begin{pmatrix}0&0&0\\
0&0&i\\0&-i&0
\end{pmatrix},\\
U^{(7)}_S&= \begin{pmatrix}1&0&0\\
0&0&0\\0&0&0
\end{pmatrix},&
U^{(8)}_S&= \begin{pmatrix}0&0&0\\
0&1&0\\0&0&0\end{pmatrix},&
U^{(9)}_S&= \begin{pmatrix}0&0&0\\
0&0&0\\0&0&1
\end{pmatrix}.
\end{aligned}
\label{F40}
\end{eqnarray}
The elements of the symbol of the density matrix \eqref{F35} that can be derived by the dequantizers~\eqref{F40} are
\begin{eqnarray}
\label{F41}
&f_{\rho,s}^{(1)}&= s_1^{(1)}=\sqrt2(p_1^{(31)}-1/2),\quad
f_{\rho,s}^{(2)}= s_2^{(1)}= \sqrt2(p_2^{(31)}-1/2),\nonumber\\
&f_{\rho,s}^{(3)}&= s_1^{(2)}=\sqrt2(p_1^{(21)}-1/2),\quad
f_{\rho,s}^{(4)}= s_2^{(2)}= \sqrt2(p_2^{(21)}-1/2),\nonumber\\
&f_{\rho,s}^{(5)}&= s_1^{(3)}=\sqrt2( p_1^{(32)}-1/2),\quad
f_{\rho,s}^{(6)}=s_2^{(3)}= \sqrt2(p_2^{(32)}-1/2),\nonumber\\
&f_{\rho,s}^{(7)}&= s_3^{(1)}+s_3^{(2)}=(p_3^{(22)}-1/2)+(p_3^{(33)}-1/2),\nonumber\\
&f_{\rho,s}^{(8)}&=1/2-s_3^{(2)},\quad
f_{\rho,s}^{(9)}=1/2-s_3^{(1)}.
\end{eqnarray}
In accordance with the reasoning described in section~\ref{sec:QbS} this symbol can be called mean value symbol. As operators \eqref{F40} satisfy the orthogonality condition \eqref{F22}, therefore the set of dequantizers \eqref{F40} is a self-dual set. Naturally, Eq.~\eqref{F2d} can be used to derive the density operator \eqref{F35} using the quantizers \eqref{F40} and the symbol \eqref{F41}.

The construction of dequantizer--quantizer formalism can be extended to an arbitrary qudit system. Following the derivation described in~\cite{CMM1, mmanko-JPCS, Milestones, Entropy-MA-2018}, matrix elements of the $d\times d$ density matrix can be expressed in terms of artificial qubit probabilities $p_{1,2,3}^{(j,k)}$ as
\begin{eqnarray}
\rho_{jk}&=&p_1^{(jk)}-\frac{1}{2}+i\left(p_2^{(jk)}-\frac{1}{2}\right), \qquad j>k,\nonumber\\
\rho_{jj}&=&1-p_3^{(jj)}, \qquad j>1,\\
\rho_{11}&=&\sum_{j=2}^d p_3^{(jj)}-d+2.\nonumber
\end{eqnarray}
In the above equation $p_{1,2}^{(jk)}$ ($j,k=1,2,\dots,d$, $j>k$) and $p_3^{(jj)}$ ($j=2,\dots,d$) are the probabilities of spin-$1/2$ projections equal to $+1/2$ on axes $x$, $y$, $z$, respectively, measured on $\binom{d}{2}$ artificial qubits composed of the states $\ket{j}$ and $\ket{k}$ in the same logic as described for qutrits.
The given probability representation of the qudit state can be used to construct two sets of $d^2$ quantizer--dequantizer pairs using the procedure described for qubits and qutrits.
The first set of dequantizers provides the symbol of the density operator expressed by the probabilities $p_{1,2,3}^{(j,k)}$ while the second set of dequantizers leads to the symbol of the density operator expressed in terms of mean values of spin projections.
For finding the matrix elements of the dequantizer operators for either of the two representations of qudits the system of linear equations containing $d^4$ equations derived from Eq.~\eqref{F1d} has to be solved that can be performed numerically for arbitrary dimension.

\section{Star product formalism for qubits}\label{sec:KASPDQD}

In this section we present the star-product formalism for the probability and mean value representations of qubits. If the symbols $f_A$ and $f_B$ of two arbitrary operators $\hat{A}$ and $\hat{B}$ defined in the Hilbert space of qubits are known then the symbol of the product of the two operators can be calculated as the star product of the two symbols defined as
\begin{eqnarray}
\label{F45}
(f_A*f_B)^{(k)}&=&\sum_{m,n} f_A^{(m)}f_B^{(n)}K_{m,n}^k,\\
\nonumber
K_{mn}^k&=&\Tr\left( D^{(m)} D^{(n)} U ^{(k)}\right);\;\;m,n,k=1,2,3,4.
\end{eqnarray}
Let us find now the values $K_{mn}^k$ for the sets of dequantizers \eqref{F10} and quantizers \eqref{F13}. We present them in the form of four matrices $K^k=\|K_{mn}^k\|$, $k,m,n=1,2,3,4$
\begin{eqnarray}
\label{F49}
\nonumber
K^1&=&\frac12\begin{pmatrix}
2&0&0&0\\0&2&-1+i&-1-i\\
0&-1-i&1&i\\0&-1+i&-i&1
\end{pmatrix},\\
\nonumber
K^2&=&\frac12\begin{pmatrix}
2&0&-1-i&-1+i\\0&2&0&0\\
-1+i&0&1&-i\\-1-i&0&i&1
\end{pmatrix},\\
\nonumber
K^3&=&\frac12\begin{pmatrix}
2&2i&-1-i&-1-i\\-2i&2&-1+i&-1+i\\
-1+i&-1-i&3&1\\-1+i&-1-i&1&1
\end{pmatrix},\\
K^4&=&\frac12\begin{pmatrix}
2&-2i&-1+i&-1+i\\ 2i&2&-1-i&-1-i\\
-1-i&-1+i&1&1\\-1-i&-1+i&1&3
\end{pmatrix}.
\end{eqnarray}

In order to show the working of this star product let us consider an example. Operator $\hat{A}$ is considered to be the well-known Hadamard operator $\hat{H}$ represented by the matrix
\begin{equation}
{H}=\frac{1}{\sqrt{2}}\begin{pmatrix}
 1& 1\\
1 & -1
\end{pmatrix}
\end{equation}
and operator $\hat{B}$ is the density operator $\hat{\rho}$ defined in Eq.~\eqref{F15}. Using Eq.~\eqref{F1} and the dequantizers defined in Eq.~\eqref{F10} one can find the elements of the probability symbol of the Hadamard operator as
\begin{equation}
\label{ExfH}
\begin{aligned}
f_{H1}&=\frac{1}{\sqrt{2}},\qquad&
f_{H2}&=0,\\
f_{H3}&=\frac{1}{\sqrt{2}},\qquad&
f_{H4}&=-\frac{1}{\sqrt{2}}.
\end{aligned}
\end{equation}
The symbol of the product of the operators $\hat{H}$ and $\hat{\rho}$  can be determined using Eq.~\eqref{F45} by substituting Eqs.~\eqref{F18}, \eqref{F49} and \eqref{ExfH} resulting in
\begin{eqnarray}
\label{ExfHrho}
\begin{aligned}
f_{H\rho,1}&=\frac{2p_3 - 2ip_2 + 2p_1 + i - 1}{2^{\frac{3}{2}}},\\
f_{H\rho,2}&=\frac{(1-i)p_3+(1+i)p_1-1}{\sqrt{2}},\\
f_{H\rho,3}&=\frac{2p_3 + 2ip_2 + 2p_1 - i - 1}{2^{\frac{3}{2}}},\\
f_{H\rho,4}&=\frac{2p_3 - 2ip_2 + 2p_1+i-3}{2^{\frac{3}{2}}}.
\end{aligned}
\end{eqnarray}
We note that this symbol can be derived by applying Eq.~\eqref{F1d} to the product of operators $\hat{H}$ and $\hat{\rho}$.

An interesting application of the kernels defined in Eq.~\eqref{F45} is that one can derive the structure constants of the Lie algebra of the quantizer operators with the help of these kernels. The structure constants $C_{mn}^k$ define the commutation relations $[L_m,L_n]=\sum_kC_{mn}^kL_k$ of Lie algebra generators $L_k$.

Let us consider the difference of two kernels
\begin{eqnarray}
\label{F46}
K_{mn}^k-K_{nm}^k&=&\Tr\left( D^{(m)} D^{(n)} U^{(k)}\right)-\Tr\left( D^{(n)} D^{(m)} U^{(k)}\right)\nonumber\\
&=&\Tr\left( ( D^{(m)} D^{(n)}- D^{(n)} D^{(m)}) U^{(k)}\right)\nonumber\\
&=&\Tr\left(\sum_{l=1}^4C_{mn}^l D^{(l)} U^{(k)}\right)=
C_{mn}^k.
\end{eqnarray}
We see that the difference \eqref{F46} of two kernels is a structure constant of the Lie algebra formed by quantizers \eqref{F13}.

Let us find now the values $K_{mn}^k$ for the set of self-dual dequantizers \eqref{F21} of the mean value representation. Again, we present them in the form of four matrices $K_S^k=\|K_{mn}^k\|$, $k,m,n=1,2,3,4$
\begin{eqnarray}
\label{F51}
\begin{aligned}
K^1_S &=\frac{1}{\sqrt{2}}\begin{pmatrix}
0 & 0 & 0 & 1\\
0 & 0 & i & 0\\
0 & -i & 0 & 0\\
1 & 0 & 0 & 0
\end{pmatrix},\qquad&
K^2_S &=\frac{1}{\sqrt{2}}\begin{pmatrix}
0 & 0 & -i & 0\\
0 & 0 & 0 & 1\\
i & 0 & 0 & 0\\
0 & 1 & 0 & 0
\end{pmatrix},\\
K^3_S &=\frac{1}{\sqrt{2}}\begin{pmatrix}
0 & i & 0 & 0\\
-i & 0 & 0 & 0\\
0 & 0 & 0 & 1\\
0 & 0 & 1 & 0
\end{pmatrix},\qquad&
K^4_S &=\frac{1}{\sqrt{2}}\begin{pmatrix}
1 & 0 & 0 & 0\\
0 & 1 & 0 & 0\\
0 & 0 & 1 & 0\\
0 & 0 & 0 & 1
\end{pmatrix}
\end{aligned}
\end{eqnarray}

In order to present the star product in action let us consider an example. Operator $\hat{A}$ is again considered to be Hadamard operator and operator $\hat{B}$ is the density operator $\hat{\rho}$ defined in Eq.~\eqref{F15}. Using Eq.~\eqref{F1} and the self-dual dequantizers defined in Eq.~\eqref{F21} one can find the probability symbols of the Hadamard operator as
\begin{equation}
\label{ExsH}
\begin{aligned}
f_{H1}&=1,\qquad&
f_{H2}&=0,\\
f_{H3}&=1,\qquad&
f_{H4}&=0.
\end{aligned}
\end{equation}
Symbols of the product of operators $\hat{H}$ and $\hat{\rho}$  can be determined using Eq.~\eqref{F45} by substituting Eqs.~\eqref{F24}, \eqref{F51} and \eqref{ExsH} resulting in
\begin{eqnarray}
\label{ExsHrho}
\begin{aligned}
s_{H\rho,1}&=\frac{1+i-2ip_2}{2},\\
s_{H\rho,2}&=ip_1-ip_3,\\
s_{H\rho,3}&=\frac{1-i+2ip_2}{2},\\
s_{H\rho,4}&=p_3+p_1-1
\end{aligned}
\end{eqnarray}
We note that this symbol can be derived by applying Eq.~\eqref{F1d} to the product of $\hat{H}$ and $\hat{\rho}$.

For self-dual systems Eq.~\eqref{F46} leads to 
\begin{eqnarray}
C_{mn}^k=K_{mn}^k-K_{nm}^k=\Tr\big( \hat U(m)\hat U(n)\hat U (k)\big)-\Tr\big( \hat U(n)\hat U(m)\hat U (k)\big),
\end{eqnarray}
and since the properties of trace lead to the equations
\begin{eqnarray}
\label{F53}
K_{mn}^k=K_{nk}^m=K_{km}^n;\;\;K_{nm}^k=K_{kn}^m=K_{mk}^n.
\end{eqnarray}
The validity of these equations can also be verified from Eq.~\eqref{F51}. Using these equations the structure constants satisfy the equations
\begin{eqnarray}
\label{F54}
C_{mn}^k=C_{nk}^m=C_{km}^n.
\end{eqnarray}
We note that for qudit systems the generalization of Eq.~\eqref{F45} leads to $d^2$ kernel matrices. Knowing the quantizers and dequantizers these $d^2\times d^2$ matrices can be easily derived. In the case of mean value representation Eq.~\eqref{F54} describing the special properties of the structure constants remain valid.

\section{Conclusions}\label{sec:conc}

We have derived the quantizers and dequantizers for the probability and mean value representations of qubit and qutrit density matrices that are necessary for developing the star product formalism. For qubits we have also determined the kernel matrices in an explicit form which are required for the derivation of the star product of the symbols.
With the help of these kernels we have described the algebras associated with dequantizer and quantizer operators and found the structure constants of these algebras.
We have discussed how the formalism can be extended to qudit states. We have presented examples for the application of the developed formalism.

\begin{acknowledgments}
This study was performed within the framework of scientific collaboration between the Lebedev Physical Institute of the Russian Academy of Sciences and the Wigner Research Centre for Physics (Project ``Quantum correlations, decoherence in the electromagnetic-field interaction with matter and tomographic approach to signal analysis.''

This research was supported by the National Research, Development and Innovation Office, Hungary (Projects No. K124351 and No. 2017-1.2.1-NKP-2017-00001 HunQuTech). The project has also been supported by the European Union (Grant Nos. EFOP-3.6.2-16-2017-00005 and EFOP-3.4.3-16-2016-00005.).
\end{acknowledgments}


\end{document}